\documentclass[12pt]{JHEP3} 
\usepackage{amsmath,amssymb,epsfig}

\def\BOX{\mathord{\vbox{\hrule\hbox{\vrule\hskip 3pt\vbox{\vskip
3pt\vskip 3pt}\hskip 3pt\vrule}\hrule}\hskip 1pt}}

\newcommand{\RR}{\mathbb{R}}

\def\pri{p_{r,i}}

\renewcommand{\Im}{\mathop{\rm Im}}

\title{Macroscopic Screening of Coulomb Potentials From UV/IR-Mixing}

\author{Robert C. Helling$^{ab}$, Jiangyang You$^b$\\
$^a$Ludwig-Maximilians-Universit\"at M\"unchen\\
Theresienstra\ss e 37\\
80333 M\"unchen, Germany\\
\\
$^b$School of Engineering and Science\\
Jacobs University\\
Postfach 750561\\
28725 Bremen, Germany\\
{\tt E-mail: helling@atdotde.de, j.you@iu-bremen.de}}
\preprint{LMU-ASC 48/07\\IUB-TH-077}
\abstract{We compute the static potential in a non-commutative theory
  including a term due to UV/IR-mixing. As a result, the potential
  decays exponentially fast with distance rather than like a power law
  Coulomb type potential due to the exchange of massless particles.
  This shows that when quantum effects are taken into account the
  introduction of non-commutativity not only modifies physics at short
  distances but has dramatic macroscopic consequences as well. As a
  result, we give a {\em lower} bound on the scale of non-commutativity
  (if present at all) to be compatible with observations.}

\begin{document}
\preprint{}



\section{Introduction}
\setcounter{equation}{0}
We have two extremely successful approaches to understand the
fundamental interactions: Quantum field theory exemplified in the
standard model of particle physics and general relativity governing
the gravitational forces dominating the universe at large scales. It
is widely believed that for a unification of these two approaches one
has to develop a description of a quantum space-time which has to
approximate the structure of a differentiable manifold at distances
larger than the Planck scale.

Probably the most conservative approach to such a quantum space-time
is to promote the coordinates to quantum operators which no longer
commute and thereby induce a fundamental quantum uncertainty of
geometry at the Planck
scale\cite{Snyder:1946qz}\cite{Doplicher:1994tu}
.
This scenario gained a lot of momentum a few years ago when it became
clear that it arises naturally from string theory in backgrounds with
non-vanishing Kalb-Ramond 2-form $B_{\mu\nu}$
\cite{Douglas:1997fm,Seiberg:1999vs,Douglas:2001ba}
. 

The simplest version of space-time non-commutativity is obtained from
commutation relations 
\begin{equation*}
  [x^\mu,x^\nu]=i\theta^{\mu\nu}
\end{equation*}
for a constant, anti-symmetric matrix $\theta$. It leads to the
Moyal-Weyl *-product for fields living in this space time
\begin{equation*}
  (f*g)(x) =  e^{\frac i2 \theta^{\mu\nu}{\frac{\partial}{\partial
y^\mu}}{\frac{\partial}{\partial z^\nu}}}f(y)g(z)\bigg|_{x=y=z}
\end{equation*}
By replacing the ordinary commutative product between fields in the
action many field theories can adapted to this non-commutative
space-time. For the standard model, this has been done in
\cite{Calmet:2001na}. In \cite{Girotti:2001dh} effective action of the
non-commutative Wess-Zumino was considered; as a result of
supersymmetry this does not lead to the UV/IR-mixing terms that we
study in this paper.

In these theories, $\theta$ plays the same role as $\hbar$ in quantum
mechanics. In quantum mechanics, the correspondence principle states
that the quantum deformation of the classical theory is only noticeable
at small scales where variations of the classical action are of the
order of $\hbar$. Similarly, one could hope that a space-time
non-commutativity with the entries of $\theta$ being of the order of
$\ell_P^2$ would only be relevant for physics at Planckian distances
or with typical momenta of the order of the Planck mass while
macroscopic physics would be indistinguishable from a theory on a
commutative space.

This expectation was shaken by \cite{Minwalla:1999px} where loop
corrections to non-commutative scalar field theory were computed.
There it was found that while planar diagrams receive no modifications
due to non-commutativity, loop momentum integrals of non-planar
diagrams have additional phase factors which generically render
integrals which are UV divergent in the commutative theory finite.
Thus the non-commutativity acts as a UV regulator.

For special external momenta, and most importantly $p\to 0$, the
regulator is not effective and the UV divergence reappears now in the
form of a IR divergence. Quadratic divergences of the commutative
theory thus lead to terms of the form
\begin{equation}
  \label{eq:counter}
  \phi\frac 1{p\circ p}\phi = \phi\frac
  1{p^\mu\theta_{\mu\nu}\theta^{\nu\rho}p_\rho}\phi 
\end{equation}
in the effective action. These terms notably are not of the form of
the *-product interactions of the classical theory and it seems likely
that at each loop order new counter terms are needed. This would
render the non-commutative theories non-renormalisable. 

This expectation turned out to be too pessimistic as in
\cite{Grosse:2005da,Grosse:2004ik,Grosse:2004yu} it was shown that
$\phi^4$ theory on non-commutative $\RR^4$ is in fact renormalisable.
There it was shown that only a single further counter-term is
required. In those papers, for technical reasons this counter-term was
taken to be
\begin{equation}
  \label{eq:grosseterm}
  x^\mu*\phi*\phi*x_\mu
\end{equation}
to make the action symmetric under Fourier transform. It cancels the
divergence of the form (\ref{eq:counter}) as it has the same scaling
dimension and $\theta_{\nu\mu}x^\nu*$ acts like $\partial_\mu$. Thus,
up to finite terms, it is highly plausible\cite{Grosse}\ that
including a counter term of the form of (\ref{eq:counter}) also yields
a renormalisable $\phi^4$ theory although an actual proof seems to be
much more involved and we do not attempt it here. For the case of
self-dual theta this theory has in the meantime been shown to be
renormalisable\cite{Gurau:2008vd}. This motivates the
approach we will take in this paper: The term (\ref{eq:counter}) is
not just the result of a one loop calculation but it should be thought
of as part of the quadratic part of the effective action which if not
present right from the start is generated by quantum corrections. Its
contribution can be resummed by instead of the usual Feynman
propagator using the functional inverse of $\BOX+\frac 1{p\circ p}$ as
the propagator.

Thus we take our theory to be given by
\begin{equation}
  \label{eq:action}
  {\cal S}=\int\!\! d^4x\, \bigg(\partial^\mu\phi\partial_\nu\phi+\phi\frac
  1{p\circ p}\phi+\frac\lambda{4!}\phi*\phi*\phi*\phi\bigg)
\end{equation}
where as usual $p_\mu=i\partial_\mu$. Note well that this new term is
non-perturbative in $\theta$ and becomes singular in the commutative
limit. For simplicity of notation we have not included a mass term as
this does not alter the conclusions of our argument. In addition, we
could have included a coupling constant in front of the new term but
as in the following we are dealing just with the quadratic part of the
theory (and will not need the $\phi^4$-interaction once it generated
the new term), such a coupling constant can always be absorbed in the
definition of $\theta$.

The choice of considering a scalar field is merely for clarity of
presentation. The discussion of this paper directly generalizes to any
other field theory with UV/IR-mixing. This justifies our misnomer
``Coulomb potential'' for a $1/r$ potential transmitted by a massless
scalar instead of a vector field.

In the existing literature, an effective action including this one
loop induced term has been discussed as giving a potential in momentum
space which has its minimum away from $p_\mu=0$ thus leading to a
spontaneous breaking of translation invariance and the formation of
stripes%
\cite{VanRaamsdonk:2001jd,Armoni:2001uw,Guralnik:2002ru,Castorina:2003zv}.
Lattice calculations have confirmed that this effect persists
non-perturbatively\cite{Bietenholz:2003hx,Bietenholz:2006cz} (and
\cite{Panero:2006cs,Panero:2006bx} in the case of the fuzzy sphere).

In this paper however, we take a different point of view and treat all
terms quadratic in the field as part of the free theory determining
the propagator. We will find that instead of the power law behavior
of the propagator as $1/p^{D-2}$ in the commutative theory the
non-commutative propagator is decaying exponentially fast. Consequently,
the force the field is transmitting is short ranged and effectively
cut-off at the scale of the non-commutativity. We find that even a
small amount of non-commutativity has drastic effects at macroscopic
scales as it screens long range forces. Our conclusion will thus be
that it is not correct to think of non-commutativity modifying a
theory only at short scales --- possibly the Planck scale. As soon as
quantum effects which result in UV/IR mixing are considered, the theory
is drastically changed at scales large compared to the
non-commutativity.

\section{Screened Coulomb interaction}
For concreteness, let us work in 3+1 dimensional Minkowski space. We
want to compute the potential due to a source 
\begin{equation*}
  J(x) = \int \!\!\! d\tau\, \delta^{(4)}(x-r(\tau)).
\end{equation*}
Let us assume the source is static and the force is mediated by a
field with momentum space propagator $G(p)=G(E,\vec p)$. Then the
potential is given by
\begin{equation*}
  V(t,\vec x) = \int\!\!\! d^4x'\int \!\!\! d^4p\, G(p)e^{ip(x-x')}J(x')
  = \int\!\!\! d^3\vec p\, G(0,\vec p)e^{i\vec p\cdot (\vec x-\vec r)}.
\end{equation*}
Thus, in order to obtain the potential we have to compute the spatial
Fourier transform of the propagator at $E=0$. 

For the scalar theory with the action (\ref{eq:action}), the momentum
space propagator is given by
\begin{equation*}
  G(p) = \frac 1{p^2+\frac 1{p\circ p}}.
\end{equation*}
Any finite coefficient of the $p\circ p$-term in the action can be
absorbed in the definition of $\theta^{\mu\nu}$ as in this paper we
will only be concerned with the propagator and the only appearance of
$\theta^{\mu\nu}$ is quadratically in $\circ$.

As a warm up, we will consider a Euclidean theory with self-dual
\begin{equation*}
  \theta^{\mu\nu} =
  \begin{pmatrix}
    0&\theta&0&0\\
    -\theta&0&0&0\\
    0&0&0&\theta\\
    0&0&-\theta&0
  \end{pmatrix},
\end{equation*}
as in this case $p\circ p=\theta^2p^2$ and the propagator only depends
on $p^2$. We can easily evaluate the static potential in polar
coordinates:
\begin{equation}
  \label{eq:sdpot}
  V(r)= \int \!\!\! d^3\vec p \,\frac{e^{i\vec p\cdot \vec r}}{\vec
    p^2+\frac 1{\theta^2\vec p^2}} = 2\pi^2 \frac{e^{-\frac
      r{\sqrt{2|\theta|}}} \cos\left(\frac r{\sqrt{2|\theta|}}\right)} r
\end{equation}
The effect of the non-commutativity is to dress the $1/r$ Coulomb
potential with an oscillatory factor and an exponential decay over the
scale of the non-commutativity. The power law character of the force
law has been replaced by a finite interaction range governed by the
length scale of the non-commutativity. Indeed, the deviation from the
Coulomb behavior is most pronounced at macroscopic distances $r\gg
\sqrt{|\theta|}$ and we have to conclude that it is incorrect to think
of the non-commutativity as a correction affecting physics only at
very short distances.

It might come as a surprise that the commutative $1/r$ potential is
not recovered as a smooth $\theta\to 0$ limit but in the opposite
$\theta\to\infty$ limit. This however is due to the nature of the
UV/IR-mixing term which blows up for small $\theta$. 

Our result demonstrates that it is essential to fully resum the
contribution of all orders in $\theta$ in order not to miss the
effects of the non-locality of the *-product. The effect we study in
this paper would have been missed in any investigation which had
expanded the theory in $\theta$ and then had truncated the infinite
series after a finite number of terms.

\EPSFIGURE[p]{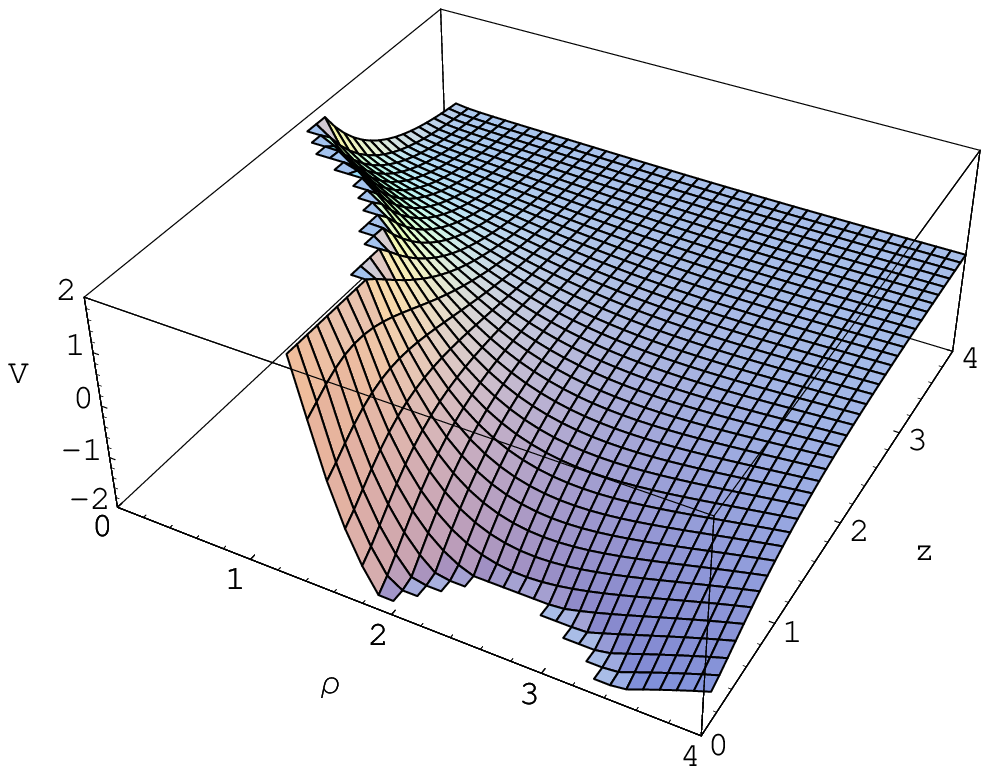,width=0.55\hsize}{The static potential,
coordinates in units of $\sqrt{\theta}$.}
Let us now consider a more general non-commutativity. We want to
consider a situation with (possibly after an appropriate boost) only
spatial non-commutativity. We can then alsways pick coordinates such that
\begin{equation*}
  \theta^{\mu\nu} =
  \begin{pmatrix}
    0&0&0&0\\
    0&0&\theta&0\\
    0&-\theta&0&0\\
    0&0&0&0
  \end{pmatrix},
\end{equation*}
the non-commutativity is only in the $xy$-plane and $\theta>0$. In cylindrical
coordinates, after a change of coordinates $p\mapsto p/\sqrt{\theta}$,
we have to evaluate
\begin{equation*}
  V(\varrho,z)= \frac 1{\sqrt{\theta}}\int_0^\infty\!\!\! p_r\,
  dp_r\int_0^{2\pi}\!\!\! dp_\phi \int_{-\infty}^\infty\!\!\! dp_z\, \frac
  {e^{i(p_r \varrho \cos p_\phi+p_z z)/\sqrt{\theta}}}{p_z^2+p_r^2+\frac 1{p_r^2}} 
\end{equation*}
Figure 1 shows this static potential evaluated numerically. One can see
it is indeed decaying exponentially fast.

The $p_\phi$ and $p_z$ integrals are easily evaluated and we find
\begin{equation*}
  V(\varrho,z) =\frac {4\pi^2}{\sqrt{\theta}} \int_0^\infty\!\!\!dp_r\, p_r
  \frac {e^{-|z|\sqrt{p_r^2+\frac
        1{p_r^2}}\big /\sqrt{\theta}}J_0(p_r \varrho/\sqrt{\theta})}
  {\sqrt{p_r^2+\frac 1{p_r^2}}}.
\end{equation*}
We cannot evaluate this remaining integral analytically. However, it
is sufficient to determine it asymptotically for distances
$\sqrt{\varrho^2+z^2}$ large compared to $\sqrt\theta$ using the
method of steepest descent. We fix the direction $c=\varrho/|z|$ use
the asymptotic form of the Bessel function
\begin{equation*}
  J_0(x) \to \sqrt{\frac 2{\pi x}} \cos(x-\pi/4).
\end{equation*}
Eventually
\begin{equation}
  \label{eq:vint}
  V(cz,z)\to \sqrt{\frac{32\pi^3}{\sqrt\theta c|z|}}\int_0^\infty\!\!\!
  dp_r\, \sqrt{\frac{p_r}{p_r^2+\frac 1{p_r^2}}} e^{-|z|\sqrt{p_r^2+\frac
        1{p_r^2}}\big /\sqrt{\theta}}\cos\left(\frac{p_r
        c|z|}{\sqrt{\theta}}-\frac\pi 4\right).
\end{equation}
Next, we rewrite the cosine as a sum of two complex exponentials. 
For $|z|\gg\sqrt\theta$ the integral is dominated by the
contribution from the points where the phase $\Phi(p_r) =
\sqrt{p_r^2+\frac 1{p_r^2}}+ic\alpha p_r$ is stationary (with
  $\alpha=\pm 1$). For each $\alpha$ there are
contributions from four stationary points $p_{r,i}$ (which are a
function of $c$ only but not of $z$ or $\theta$) obeying
\begin{equation}
  \label{eq:stationary}
  p_{r,i} -\frac 1{p_{r,i}^3} = -i\alpha c\sqrt{p_{r,i}^2+\frac 1{p_{r,i}^2}}.
\end{equation}
By evaluating the $p_r$ integral to leading order in $z/\sqrt{\theta}$ we find
\begin{equation}
  \label{eq:Vsum}
  V(z,cz)= \sum_{\pm,i} \frac {4\pi^2}{|z|} \left(1+\frac
  1{\pri^4}\right)^{\frac 14}\frac{\pm\pri^4}{\sqrt{c(2+6\pri^4)}}
  e^{\pm{(\Phi(p_{r,i})|z|/\sqrt\theta)-\pi i/4}}\Big(1+O\big(\frac 1{|z|}\big)\Big) 
\end{equation}
as explained in the appendix. As detailed there, the sum has to be
taken over all solutions $\pri$ of 
\begin{equation}
  \label{eq:ps}
  \frac 1{\pri^4} = 1-\frac {c^2\pm c\sqrt{c^2-8}}2.
\end{equation}
By analyzing the integrand of (\ref{eq:vint}) for large and small
$p_r$ it can be seen that $V(cz,z)\to 0$ for large $z$. Thus we can
conclude that the sign in the exponential in (\ref{eq:Vsum}) will
always be such that the real part of the coefficient of $|z|$ is
positive. In the appendix, it is shown that this is indeed the case.

The expression (\ref{eq:Vsum}) constitutes the main result of this
work. It shows that for distances $z$ large compared to the length
scale of non-commutativity $\sqrt\theta$, the interaction which is
transmitted by the field $\phi$ dies off exponentially fast in contrast to
the commutative situation where it only falls off like $1/z$. The
effect is as if due to the non-commutativity the field $\phi$ had
acquired a direction dependent mass of the order of $1/\sqrt\theta$.

\section{Conclusions}
In this paper, we investigated the effect of non-commutativity on the
long range interactions mediated by a scalar field taking into account
UV/IR mixing.  This effect due to loop diagrams induces new quadratic
terms in the effective action with negative powers of $\theta$, the
scale of non-commutativity.

We showed that by including these terms into a resummed propagator the
power law effective potential is modified to a Yukawa type potential
with exponential fall-off with a direction dependent decay length of
the order of $\sqrt\theta$. 

This calculation shows explicitly that the assumption motivated by a
classical analysis that the deformation leading to non-commutativity
of space-time coordinates only modifies physics at short distances
(typically the Planck scale) is not true once quantum effects are
taken into account. Rather, the drastic consequences of the
non-commutativity is that it switches off the interaction for
macroscopic scales.

This raises the question of viability of simple non-commutative models
for phenomenology given that at macroscopic scales we do observe long
range power-law interactions but no signs of non-commutativity. In this
paper we have only treated a single scalar field but the argument
should trivially generalize to fields of higher spin as soon as the
theory has UV/IR-mixing. The specific term $1/p\circ p$ arises from
integrals which are quadratically divergent in the commutative
theory. Thus the argument should directly apply for example for the
Higgs self-energy in the standard model giving a lower bound on
$\sqrt\theta$ at the order of the inverse Higgs mass. Terms with
milder divergence lead to slightly different terms in the effective
action which however will have a similar behavior that they become
large for small $\sqrt\theta$ since they arise from divergences which
are regularized by the non-commutativity resulting in similar
macroscopic consequences of the non-commutativity and resulting in
simiar bounds.

Thus it is likely a generic situation that quantum effects
transmit non-localities due to non-commutativity which classically
appear at short scales to arbitrary large scales. This should be seen
as a challenge to model-building based on non-commutative space-time.

\bigskip

\goodbreak
\acknowledgments We have profited a lot from discussions with Harald
Grosse and Peter
Schupp. Both of us are grateful to Jacobs University for financial
support. RCH thanks Eli\-te\-netz\-werk Bayern for funding.
\nobreak 

\appendix
\section{The effective potential is decaying exponentially}
In this appendix, it will be shown that the real part of the phase
$\Phi(p_{r,i})$ at the stationary points is positive and thus the
effective potential in (\ref{eq:Vsum}) is indeed decaying at large
distances.

We want to determine the sign of the real part of 
\begin{equation*}
  \Phi(\pri) = \sqrt{\pri^2+\frac 1{\pri^2}}+i\alpha c \pri
\end{equation*}
at its stationary points with $\alpha=\pm 1$ and $c\ge 0$. As we
pick the branch-cut to be along the negative real
axis, all square-roots will have non-negative real parts. Thus it is
sufficient to show that at the critical points we have
$\alpha\Im(\pri)\le 0$.

Setting the derivative of $\Phi$ to zero yields
\begin{equation}
  \label{eq:statcond}
  \alpha\left(\pri -\frac 1{\pri^3}\right) = -i c\sqrt{\pri^2+\frac 1{\pri^2}}.
\end{equation}
Squaring this relation gives a quadratic equation for $1/\pri^4$ which
is solved by
\begin{equation}
  \label{eq:p4sol}
  \frac 1{\pri^4} = 1-\frac c2 \left( c+\alpha\beta\sqrt{c^2-8}\right)
\end{equation}
with another sign $\beta=\pm 1$. However, as to obtain this solution
we have squared (\ref{eq:statcond}), only four of the eight solutions for
$\pri$ in (\ref{eq:p4sol}) will really solve
(\ref{eq:statcond}). However, the other solutions will solve
(\ref{eq:statcond}) for the other choice of $\alpha$. Therefore,
instead of fixing $\alpha$ and determining which four of the eight
$\pri$ is really a solution, we find the appropriate $\alpha_i$ for
each of the four $\pri$. 

Let us first consider the case $c^2\ge 8$. From  (\ref{eq:statcond})
and the non-negative real part of square-roots we find that the
imaginary part of $\alpha_i \pri (1-1/\pri^4)$ is negative. But
(\ref{eq:p4sol}) implies that $1-1/\pri^4$ is real and positive. Thus
the imaginary part of $\alpha_i\pri$ is negative and indeed
$\exp(-|z|\Phi(\pri)/\sqrt{\theta}$ decays for large $|z|$. 

\def\pr#1{p_{r,#1}}

It remains to consider $c^2<8$. As with any solution $\pri$ of
(\ref{eq:p4sol}) also $i^n\pri$ are solutions, we can first consider the
solution such that $|\arg(\pr 1)|\le\pi/4$. In this case, it is easy to see
that the imaginary part of $\pr 1 -1/\pr 1^3$ has the same sign as the
imaginary part of $\pr 1$. Again using (\ref{eq:statcond}) and the
non-negative real part of square-roots implies that $\Im(\alpha_1\pr 1)$
is not positive. Substituting the next solution $\pr 2=-\pr 1$ in
(\ref{eq:statcond}), the RHS does not change and we have
$\alpha_2=-\alpha_1$ to compensate the change of sign of $\pri+1/\pri^3$. Thus
$\alpha_1\pr 1=\alpha_2\pr 2$.

\newcommand{\sgn}{\operatorname{{\mathrm sgn}}}

Next, consider $\pr 3=i\pr 1$ which has positive imaginary part. For
any complex number $z=x+iy$ one has
\begin{equation*}
  \Im\left(z^2+\frac 1{z^2}\right)= 2 xy\left(1-\frac 1{|z|^4}\right)
\end{equation*}
From the explicit solution we have $|1/\pri^4|= 1+c^2$. Thus $\Im(\pr
1^2+1/\pr 1^2)$ has the opposite sign of $\Im(\pr 1)$. Using this
together with $\sqrt{-z} = -i \sqrt z \sgn(\Im(z))$ in evaluating the
RHS of (\ref{eq:statcond}) for $\pr 3$ gives $\alpha_3=-1$. Thus
$\Im(\alpha_3\pr 3)\le 0$ as well. The same is true for $\pr 4=-\pr 3$
as can be shown using the same argument as in the case of $\pr 2$. 

This concludes our proof that for all stationary points $\pri$ of $\Phi$ the
sign $\alpha_i$ happens to adjust to make $\Im(\alpha_i\pri)\le 0$ and
thus the effective potential decays at large distances.

\bibliographystyle{JHEP}
\bibliography{uvir}

\end{document}